\documentclass[a4paper,USenglish]{lipics}

\usepackage{microtype}

\usepackage{xspace}

\title{Small Universal Petri Nets with Inhibitor Arcs} 

\author{Sergiu Ivanov}
\author{Elisabeth Pelz}
\author{Sergey Verlan}
\affil{Laboratoire d'Algorithmique, Complexit\' e et Logique, Universit\'e Paris Est\\
  61, av. du g\'en. de Gaulle, 94010 Cr\'eteil, France\\
  \texttt{\{sergiu.ivanov,pelz,verlan\}@u-pec.fr}}

\authorrunning{S. Ivanov, E. Pelz, S. Verlan} 

\Copyright{S. Ivanov, E. Pelz, S. Verlan}


\serieslogo{}
\volumeinfo
  {Billy Editor, Bill Editors}
  {2}
  {Conference title on which this volume is based on}
  {1}
  {1}
  {1}
\EventShortName{}
\DOI{10.4230/LIPIcs.xxx.yyy.p}

\newcommand{\C}{\ensuremath{\mathfrak C}\xspace}
\newcommand{\M}{\ensuremath{\mathcal M}\xspace}

\newcommand\NN{\mathbb{N}}
\def\rkp#1#2#3{(q_{#1},R#2P,q_{#3})}
\def\rkzm#1#2#3#4{(q_{#1}, R#2ZM,q_{#3},q_{#4})}

\usepackage{enumerate}

\newcommand{\place}[1]{\ensuremath{#1}}
\newcommand{\trans}[1]{\ensuremath{#1}}

\begin{document}

\maketitle

\begin{abstract}
We investigate the problem of construction of small-size universal Petri nets
with inhibitor arcs. We consider four descriptional complexity parameters:
the number of places, transitions, inhibitor arcs, and the maximal degree of
a transition,
each of which we try to minimize.
 We give six constructions having the following values of parameters (listed in
the above order): $(30,34,13,3)$, $(14, 31, 51, 8)$, $(11, 31, 79, 11)$,
$(21,25,13,5)$, $(67, 64, 8, 3)$, $(58, 55, 8, 5)$ that improve the few known
results on this topic. Our investigation also highlights several interesting
trade-offs.
\end{abstract}

\section{Introduction}

The concept of universality was first formulated by A. Turing
in~\cite{turing1936}. He constructed a universal (Turing) machine capable of
simulating the computation of any other (Turing) machine. This universal
machine takes as input a description of the machine to simulate, the contents
of its input tape, and computes the result of its execution on the given input.
The universal Turing machine is extremely important as it provided the first
theoretical basis for the construction of modern computers -- this becomes
obvious if we consider the latter as fixed universal devices which receive as
input the coding of the algorithm solving a problem (the program) and the input
data, and which produce the result of the execution of the algorithm (program)
on that input data.

More generally, the universality problem for a class of computing devices (or
functions) \C consists in finding a fixed element \M of \C able to simulate the
computation of any element $\M'$ of $\C$ using an appropriate fixed encoding.
More precisely, if $\M'$ computes $y$ on an input $x$ (we will write this as
$\M'(x)=y$),
then $\M'(x)=f(\M(g(\M'),h(x)))$, where $h$ and $f$ are the encoding and
decoding functions, respectively, and $g$ is the function retrieving the number
of $\M'$ in some fixed enumeration of \C. These functions should not be too
complicated, otherwise the universal machine will be trivial, e.g. when $f$ is
partially recursive the machine can contain only one instruction -- stop. It is
commonly admitted that general recursive functions can be used for encoding and
decoding. The typically used functions are $f(x)=\log_2(x)$ and $h(x)=2^x$.
The element
\M is called (weakly) universal for \C. We shall call \M strongly
universal (for \C) if the encoding and decoding functions are identities.

Some authors~\cite{Malcev,Korec} implicitly consider only the strong notion of
universality as the encoding and decoding functions can perform quite
complicated transformations, which are not necessarily doable in the original
devices. For example,
Minsky's proof
of (weak) universality of register machines
with two counters~\cite{Minsky67} makes use of exponential (resp. logarithmic)
encoding (resp. decoding) functions, while it is known that such functions
cannot be computed on these machines~\cite{Barzdin}. We refer to~\cite{Korec}
for a more detailed discussion of different variants of the universality.
Generally, the class of all partially recursive functions is considered as \C,
but it is possible to have a narrower class, e.g. the class of all primitive
recursive functions, which is known to admit a universal generally recursive
function~\cite{Malcev}.

In 1956 Shannon~\cite{Shanon} considered the question of finding the smallest
possible universal Turing machine where the size is calculated as the number of
states and symbols. In the early sixties, Minsky and Watanabe had a running
competition to see who could find the smallest universal Turing
machine~\cite{Minsky62,Watanabe61}. Later, Rogozhin showed the construction of
several small universal Turing machines~\cite{Rogozhin96}. An overview of
recent results on this topic can be found in~\cite{NW12}. Other computational
models were also considered, e.g. cellular automata~\cite{vNeumann66} with a
construction of universal cellular automata of rather small size;
see~\cite{Ollinger02,Wolfram02} for an overview.

Small universal devices have mostly theoretical importance as they
demonstrate the minimal ingredients needed to achieve a complex (universal)
computation. Their construction is a long-standing and fascinating challenge
involving a lot of interconnections between different models, constructions, and
encodings.

Turing machines and cellular automata work on strings and, in order to
represent functions, unary encoding is used. Register machines~\cite{Minsky67}
manipulate numbers directly and it was shown that three registers are
sufficient for strong universality. However, corresponding constructions are
quite big as far as the number of used rules is concerned. In 1996, Korec
constructed several small universal register machines~\cite{Korec} that made
use of a reduced number of rules. These results served as the base for the
universal constructions for multiset rewriting-based models (which are
equivalent to vector addition systems and Petri nets). In~\cite{AV11} a small
universal maximally parallel multiset rewriting system is constructed. Due to
equivalences between Petri nets and multiset rewriting systems, this result can
be seen as a universal Petri net working with
max step
%
semantics. For
the traditional class of Petri nets there were no known universality
constructions for a long time. Recently, Zaitsev has investigated the
universality of Petri nets with inhibitor arcs and priorities~\cite{Zaitsev13}
and has constructed a small universal net with 14 places and 29 transitions
(for nets without priorities the same author obtained a universal net with 500
places and 500 transitions~\cite{Zaitsev2012}). We remark that inhibitor arcs
and priorities are
equivalent extensions for Petri nets in terms of computational power,
so using both concepts
together
is not necessary for universality constructions.

In this article, we perform a systematic investigation of Petri nets with
inhibitor arcs working in the classical sequential semantics. We consider as a
measure of descriptional complexity the vector $(p,t,h,d)$, where $p$ is the
number of places, $t$ is the number of transitions, $h$ is the number of
inhibitor arcs, and $d$ is the maximal degree of a transition. We remark that
from the multiset rewriting point of view these parameters correspond to the
size of the alphabet, the number of rules, the number of forbidding conditions
and the maximal size of the rule, respectively. We construct strongly universal
deterministic Petri nets with inhibitor arcs of the following sizes:
$(30,34,13,3)$, $(14, 31, 51, 8)$, $(11, 31, 79, 11)$, $(21,25,13,5)$, $(67,
64, 8, 3)$, $(58, 55, 8, 5)$. The obtained results highlight several trade-offs
between the parameters, e.g. the construction from Section~\ref{sec:mini} shows
how the decrease in the number of inhibitor arcs translates to the increase of
the number of places and transitions, while
that in
Section~\ref{sec:minp} decreases the number of states by incrementing the
number of the inhibitor arcs and the degree of the transitions.

%
%

\section{Preliminaries}

In this section we recall some basic notions and notations used in formal
language theory, and computability theory that we will need in the rest of the
paper. For further details and information the reader is referred to
\cite{handbook}.

An alphabet is a finite non-empty set of  symbols. Given an alphabet $V$, we
designate by $ V^* $ the set of all strings over $ V $, including the empty
string, $ \lambda $.
For each $x\in V^*$ and $ a \in V $, $ |x|_a $ denotes the number of occurrences of the symbol
$ a $ in $ x $. A finite multiset over $ V $ is a mapping $X: V \longrightarrow
\mathbb{N} $, where $\mathbb{N}$ denotes the set of non-negative integers.
$X(a)$ is said to be the multiplicity of $ a $ in $ X $.

\subsection{Register machines}
\label{ssec:reg-machines}

A deterministic \emph{register machine} is defined as a 5-tuple
$M=(Q,R,q_0,q_f,P),$ where $Q$ is a set of states, $R=\{R_1,\dots{},R_k\}$ is
the set of registers, $q_0\in Q$ is the initial state, $q_f\in Q$ is the final
state and $P$ is a set of instructions (called also rules) of the following
form:

\begin{enumerate}
\item \emph{(Increment)} $(p,RiP,q)\in P$, $p,q\in Q,p\ne q, R_i\in R$ (being
    in state $p$, increment register $R_i$ and go to state $q$).
\item \emph{(Decrement)} $(p,RiM,q)\in P$, $p,q\in Q,p\ne q, R_i\in R$ (being
    in state $p$, decrement register $R_i$ and go to state $q$).
\item \emph{(Zero check)} $(p,Ri,q,s)\in P$, $p,q,s\in Q, R_i\in R$ (being in
    state $p$, go to $q$ if register $R_i$ is not zero or to $s$ otherwise).
\item \emph{(Zero test and decrement)} $(p,RiZM,q,s)\in P$, $p,q,s\in Q,
    R_i\in R$ (being in state $p$, decrement register $R_i$ and go to $q$ if
    successful or to $s$ otherwise).
\item \emph{(Stop)} $(q_f,STOP)$ (may be associated only to the final state
    $q_f$).
\end{enumerate}

We note that for each state $p$ there is only one instruction of the types
above.

For conciseness, we will often refer to a rule of type $(q_j, X, q_k)$ or
eventually $(q_j, X, q_k, q_{k'})$, just by the symbol $X$.

A configuration of a register machine is given by the $(k+1)$-tuple
$(q,n_1,\dots{},n_k)$, where $q\in Q$ and $n_i\in\NN, 1\le i\le k$, describing
the current state of the machine as well as the contents of all registers. A
transition of the register machine consists in updating/checking the value of a
register according to an instruction of one of the types above and in changing the
current state to another one. We say that the machine stops if it reaches the
state $q_f$.
We say that $M$ computes a value $y\in \mathbb{N}$ on the \emph{input}
$x_1,\dots,x_n$, $x_i\in\mathbb{N}$, $1\le i\le n\le k$, if, starting from the
initial configuration $(q_0,x_1,\dots,x_n,0,\dots{},0)$, it reaches the final
configuration $(q_f,y,0,\dots{},0)$.


It is well-known that register machines compute all partial recursive
functions and only them~\cite{Minsky67}.
Therefore, every register machine $M$ with $n$ registers can be
associated with the function it computes: an $m$-ary partial recursive
function $\Phi_M^m$, where $m\leq n$.
%
Let $\Phi_0,\Phi_1, \Phi_2,\dots,$ be a fixed
enumeration of the set
of unary partial recursive functions. Then, a register machine $M$ is said to
be {\em strongly universal} if there exists a recursive function $g$ such that
$\Phi_x(y)=\Phi_M^2(g(x),y)$ holds for all $x,y \in \mathbb N$. A register
machine $M$ is said to be {\em (weakly) universal} if there exist recursive
functions $f,g,h$ such that $\Phi_x(y)=f(\Phi_M^2(g(x),h(y)))$ holds for all
$x,y \in \mathbb N$.

We also note that the power and efficiency of a register machine $M$
depend on the set of used instructions. In~\cite{Korec} several sets
of instructions are investigated. In particular, it is shown that there are
strongly universal register machines with
22 instructions of form $RiP$ and $RiZM$. Moreover, these machines can be
effectively constructed.

\begin{figure}
\begin{center}
\includegraphics[scale=0.35]{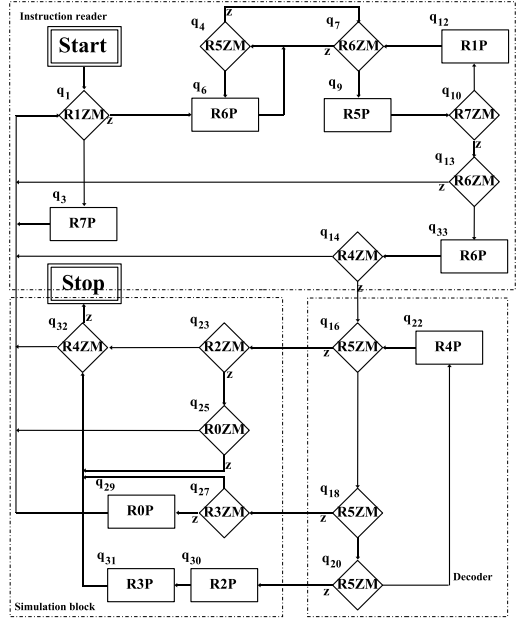}
\end{center}
\caption{Flowchart of the strongly universal machine $U_{22}$.}
\label{fig:U22}
\end{figure}

Figure~\ref{fig:U22} shows this universal register machine having 22
instructions of type $RiP$ and $RiZM$ taken from~\cite{Korec} (more precisely
in~\cite{Korec} only a machine with 32 instructions of type $RiP$, $RiM$ and
$Ri$ is constructed, and the machine below may be easily obtained from that
one by combining the instructions of type $Ri$ and $RiM$).

Here is the list of rules of this machine:
\[
\begin{array}{llll}
 \rkzm{1}{1}{3}{6} &\rkp{3}{7}{1}{} &\rkzm{4}{5}{6}{7} \\
 \rkp{6}{6}{4}{} &\rkzm{7}{6}{9}{4} &\rkp{9}{5}{10}{}\\
 \rkzm{10}{7}{12}{13} &\rkp{12}{1}{7}{} &\rkzm{13}{6}{33}{1} \\
 \rkp{33}{6}{14}{} &\rkzm{14}{4}{1}{16} &\rkzm{16}{5}{18}{23}\\
 \rkzm{18}{5}{20}{27} &\rkzm{20}{5}{22}{30} &\rkp{22}{4}{16}{}\\
 \rkzm{23}{2}{32}{25} &\rkzm{25}{0}{1}{32} &\rkzm{27}{3}{32}{1}\\
 \rkp{29}{0}{1}{} &\rkp{30}{2}{31}{} &\rkp{31}{3}{32}{}\\
 \rkzm{32}{4}{1}{f} &  (q_f, STOP)
\end{array}
\]

For the purposes of this paper, we will prefer a different representation of
register machines as graphs, the one in which the graph vertices carry only the
information about the states, while the operations and conditions are attached
to the arcs. Figure~\ref{fig:RiZMRjZM} shows such a representation of the
segment of $U_{22}$ centered at the states $q_{10}$ and $q_{13}$.  We use the
symbols $Ri$ (resp. $RiZ$) to describe the conditions in which the register
$R_i$ is required to be nonzero (resp. zero), while the symbols $RiP$ and $RiM$
stand for the operations of incrementing and decrementing $R_i$.
Such a construct is more general than a register machine, because more
than one operation or condition may be attached to arcs.  We will
later see how this generality contributes to the minimisation of
universal Petri nets.

\subsection{Petri Nets}
A \emph{Place-Transition-net} or for short, \emph{PT-net}, \emph{with inhibitor
arcs} is a construct $ N=(P,T,W,M_0)$ where $P$ is a finite set of
\emph{places}, $ T $ is a finite set of \emph{transitions}, with $ P \cap T =
\emptyset $, $ W:(P \times T) \cup (T \times P) \to \mathbb{N}\cup\{-1\} $ is
the \emph{weight function} and $ M_0 $ is a multiset over $ P $ called the
\emph{initial marking}.

PT-nets are usually represented by diagrams where places are drawn as circles,
transitions are drawn as squares annotated with their location, and a directed
arc $ (x,y) $ is added between $ x $ and $ y $ if $ W(x,y) \geq 1$. These arcs
are then annotated with their weight if this one is $ 2 $ or more. Arcs having
the weight -1 are called \emph{inhibitor} arcs and are drawn such that the arcs
end with a small circle on the side of the transition.

Given a PT-net $ N $, the \emph{pre-} and \emph{post-multiset} of a transition
$ t $ are respectively the multiset $ pre_{N(t)} $ and the multiset $
post_{N(t)} $ such that, for all $ p \in P $, for which $W(p,t)\geq 0$,
$pre_{N(t)}(p) = W(p,t) $ and  $post_{N(t)}(p) = W(t,p) $. A state of $ N $,
which is called a \emph{marking}, is a multiset $M$ over $ P $; in particular,
for every $ p \in P $, $M(p)$ represents the number of \emph{tokens} present
inside place $ p $. A transition $ t $ is \emph{enabled} at a marking $ M $ if
the multiset $ pre_{N(t)} $ is contained in the multiset $ M $ and all
inhibitor places $p$ (such that $W(p,t)=-1$) are empty. An enabled transition $
t $ at marking $ M $ can \emph{fire} and produce a new marking $ M' $ such that
$ M'=M - pre_{N(t)} + post_{N(t)} $ (i.e., for every place $ p \in P $, the
firing transition $ t $ consumes $pre_{N(t)}(p)$ tokens and produces
$post_{N(t)}(p) $ tokens). We denote this as $M\overset{t}{\longrightarrow}
M'$.

For the purposes of this paper, we have to define which kind of PT-nets can
execute computations (e.g. compute partially recursive functions). In such a
net some distinguished places $i_1,\dots,i_k$, $k>0$ from $P$ are called
\emph{input} places (which are normally different from the places marked in
$M_0$ containing the control tokens) and one other, $i_0\in P$, is called the
\emph{output} place. The computation of the net $N$ on the input vector
$(n_1,\dots{},n_k)$ starts with the initial marking $M_0'$ such that $M_0'(i_j)
= n_j$ and $M_0'(x)=M_0(x),$ for all $x\ne i_j$, $1\le j\le k$. This net will
evolve by firing transitions until deadlock in some marking $M_f$, i.e. in
$M_f$ no transition is enabled. Thus we have $M_0'\overset{*}{\longrightarrow}
M_f$ and there are no $M_f'$ and $t\in T$ such that
$M_f\overset{t}{\longrightarrow} M_f'$. The result of the computation of $N$ on
the vector $(n_1,\dots{},n_k)$, denoted by $\Phi_N^k(n_1,\dots{},n_k)$, is
defined as $M_f(i_0)$, i.e. the number of tokens in place $(i_0)$ in the final
state. Since in the general case Petri nets are non-deterministic, the function
$\Phi_N^k$ could compute a set of numbers. In this paper we consider only those
nets, which compute a unique result and which are deterministic, that
is, for any reachable marking $M$, there is at most one marking $M'$
such that $M\to M'$.  This corresponds to
labelled deterministic Petri nets in which all transitions
are labelled with the same symbol \cite{detpelz}.

\section{Universal Net with Small Transition Degree}\label{sec:mind}
In this section we will show how to construct a universal Petri net
with transitions of degree of most 3, based on the universal register
machine $U_{22}$ shown in Subsection~\ref{ssec:reg-machines}.  We will then evaluate
some basic parameters of the obtained Petri net: the number of places,
the number of transitions, the number of inhibitor arcs, and the
maximal degree of a transition.

\begin{figure}
  \centering
  \subfloat[a][$(q_j, RiP, q_k)$]{
    \includegraphics[scale=0.5]{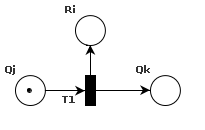}
    \label{fig:RiP}
  }
  \subfloat[b][$(q_j, RiZM, q_k, q_{k'})$]{
    \includegraphics[scale=0.5]{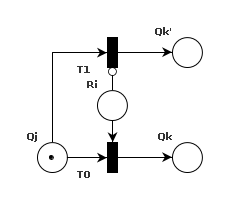}
    \label{fig:RiZM}
  }
  \caption{Petri nets simulating instructions of type $RiP$ and $RiZM$.}
\end{figure}
In $U_{22}$, only rules of type $RiZM$, and $RiP$ are used.  We will now show
how these types of rules can be implemented in Petri nets.  The general idea is
to have one place \place{Qj} for each state $q_j$ and one place \place{Ri} per
register $R_i$.  A token in place \place{Qj} means that the simulated machine
is in state $q_j$, while the number of tokens in place \place{Ri} corresponds
to the value of the register $R_i$ of the simulated machine.

A Petri net
simulating an instruction of type $RiP$ is shown in Figure~\ref{fig:RiP}.  It is rather
clear that the only transition of this Petri net correctly moves the state
token from the \place{Qj} to \place{Qk} and adds one to the value of the
register \place{Ri}.

A Petri net simulating an instruction of type $RiZM$ is shown in Figure~\ref{fig:RiZM}.
There are two transitions going out
of \place{Qj}: one moves the token from \place{Qj} into \place{Qk} and can only
be fired when \place{Ri} contains at least one token so that we can decrement
its value.  The other transition on the contrary can only be fired when
\place{Ri} contains no tokens and moves the token from \place{Qj} to
\place{Qk'}.

We can now directly construct the universal Petri net simulating the register
machine $U_{22}$ by iteratively translating all instructions. For reasons of
readability, we will refrain from showing the resulting universal Petri net.
At the initial marking this net will have one token in place \place{Q1}
corresponding to state $q_1$ of $U_{22}$.
This Petri net has 22 places for states and 8 places for registers, 30 states
all in all. There are 34 transitions in this net, as many as there are arcs in
the graph representation of $U_{22}$. Further, the number of inhibitor arcs in
this Petri net equals to the number of $RiZM$ instructions in the simulated
register machine: 13. Finally, since we only simulate rules of type $RiP$ and
$RiZM$, the maximal transition degree is 3. Obviously,
with such an approach, this is
the absolute minimal value, because there must be transitions which move the
state token between two state places and also modify a register place.
Therefore, the descriptional complexity of this net is $(30, 34, 13, 3)$.
We will refer to this net as $N_1$.

We remark that register machine $U_{22}$ supposes that the code of the machine
to be simulated is initially placed in register $R_1$ and the initial value in
register $R_2$. Under these conditions the result can be read in register $R_0$
when the machine halts. Hence, $N_1$ has two input places $R1$ and $R2$ and an
output place $R0$. Now in order to simulate an arbitrary Petri net $N$ with one
input place, $N_1$ shall be provided by the appropriate coding of $N$ and its
input in places $R1$ and $R2$ respectively. The net $N_1$ is strongly universal
because of the relation $\Phi_N(x)=\Phi^2_{N_1}(g(N),x)$, where $g$ is an
enumeration function. We remark that if we would like to obtain the complete
final marking of $N$ as a result, then this could be done by constructing
another net $N'$ which additionally encodes the final configuration of $N$ into
a number and then using a decoding function to transform it to a vector.

\section{Universal Nets with a Small Number of Places}\label{sec:minp}
In this section we construct two universal Petri nets with inhibitor arcs where
we focus on reducing the number of places. This reduction can be achieved using
two independent ideas: (a) performing several register machine instructions in
one Petri net transition and (b) using a binary encoding of the states. In both
cases the states corresponding to the states of the register machine are
reduced; in case (a) unused states are eliminated, in case (b) their number
becomes logarithmic with respect to the initial amount.

\subsection{State Compression}\label{ssec:state compression}
\begin{figure}
  \centering
  \includegraphics[scale=0.43]{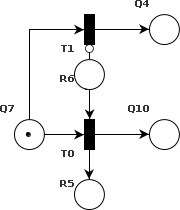}
  \caption{A Petri net simulating the states $q_7$ and $q_9$
    of $U_{22}$.}
  \label{fig:RiZMRjP}
\end{figure}

The main idea behind this optimization is that, in Petri nets, more than one
register machine operation can be performed in one transition.  For example,
the sequence of operations performed in states $q_7$ and $q_9$ can be
implemented in Petri nets as shown if Figure~\ref{fig:RiZMRjP}. Generally, we
can save one place on a $RiP$ instruction by carrying the corresponding
increment of the register
 in the preceding transition.

\begin{figure}[h]
  \centering
  \subfloat[a][An $RiZM$ followed\\ by an $RjZM$]{
    \centering
    {
    \includegraphics[scale=0.4]{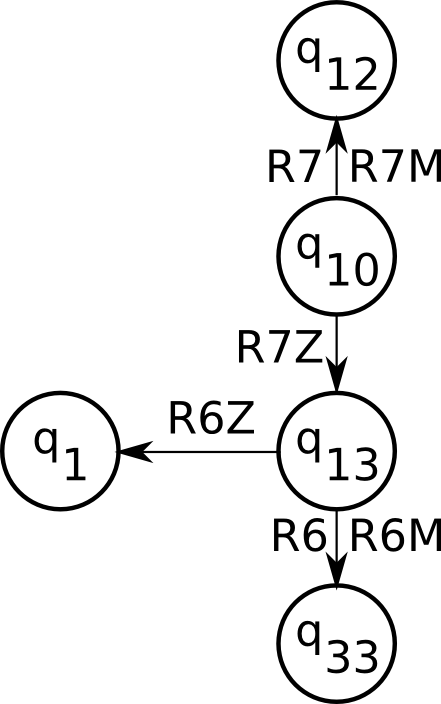}
    \hspace*{10mm}
    }
    \label{fig:RiZMRjZM}
  }
  \subfloat[a][The translation into Petri nets]{
    \centering
    {
    \includegraphics[scale=0.5]{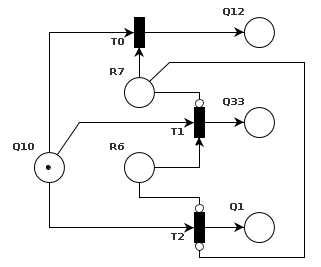}
    \hspace*{10mm}
    }
    \label{fig:RiZMRjZMPetri}
  }
  \subfloat[a][The compressed state graph]{
    \centering
    \includegraphics[scale=0.4]{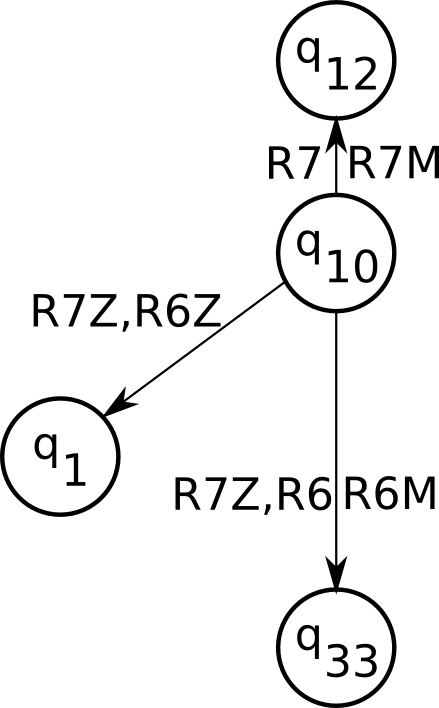}
    \label{fig:RiZMRjZMComp}
  }
  \caption{The translation of successive $RiZM$ and $RjZM$ instructions to Petri nets.}
\end{figure}

Moreover, it is possible to save a state place on two successive $RiZM$
instructions. Consider the situation depicted in Figure~\ref{fig:RiZMRjZM}.
Instead of directly translating these two instructions into Petri nets using
the patterns we have seen in the previous section, we will
merge the two potential transitions into one and thus
save a state place, as shown in Figure~\ref{fig:RiZMRjZMPetri}.

The central question now is when such an optimization technique {\em
  cannot} be applied.  The answer arises from the semantics of Petri
net transitions: before a transition is fired, the conditions are
checked whether it can be fired or not.  Therefore, in a single Petri
net transition, checking a condition on a register can only be
performed before the value of this register is modified.

Note that using a $RiZM$ instruction to check
the value of a register immediately after it was modified by a $RiP$
instruction is redundant because we know for sure that the register is not
zero.  This allows us to state that, in a single transition, we can perform any
series of actions which do not include decrements of the register $R_i$ and
{\em then} check its value.

We can now turn back to graphs of register machines and formulate the algorithm
of reducing (compressing) the number of states.  We first define the notion of
a compressible state: a state $q$ is {\em compressible} if (a) no arc leaving
$q$ checks a register modified by an $RiM$ instruction of an incoming arc, and
(b) if $q$ has no loop arcs (i.e., arcs which do not move the machine away from
state $q$). Reducing a compressible state $q$ is generally done in the
following way: for every pair of states $q_{pred}$ and $q_{succ}$ for which
there exist the arcs $q_{pred}\to q$ and $q\to q_{succ}$, we add a new arc
$q_{pred}\to q_{succ}$ which combines the conditions and operations of these
two arcs.  We then remove the state $q$ and all associated arcs.

An important remark is due here: if the arc $q_{pred}\to
q$ increments a register that is subsequently required to be nonzero by the
arc $q\to q_{succ}$, then we have to remove the (redundant) check from
the conditions of the new arc $q_{pred}\to q_{succ}$.  If, on the other
hand, this register is required to be zero, we just do not add the arc
$q_{pred}\to q_{succ}$ altogether, because such a state transition is impossible.

We also need to carefully handle the cases in which the incoming and
the outgoing arcs impose conditions on the same register.  Consider,
for example, a compressible state $q_k$, an arc $q_j\to q_k$ which
requires $R_i$ to be zero, and then another arc $q_k\to q_m$, which,
as well, requires $R_i$ to be zero.  The new arc $q_j\to q_m$ can only
be added in case $q_j\to q_k$ does {\em not} increment $R_i$, because,
obviously, the condition that $R_i$ should be zero imposed by $q_k\to
q_m$ is rendered impossible.  Generally, the following four scenarios
have to be kept in mind:
\begin{enumerate}
\item both $q_{pred}\to q$ and $q\to q_{succ}$ require $R_i$ to be zero,
\item $q_{pred}\to q$ requires $R_i$ to be zero, while $q\to q_{succ}$
    requires it to be nonzero,
\item $q_{pred}\to q$ requires $R_i$ to be nonzero, while $q\to q_{succ}$
    requires it to be zero, and
\item both $q_{pred}\to q$ and $q\to q_{succ}$ require $R_i$ to be nonzero.
\end{enumerate}

In the first situation we can only construct the new arc $q_{pred}\to
q_{succ}$ if $q_{pred}\to q$ does not increment $R_i$. In the second
case, we can only add the new arc if $q_{pred}\to q$ {\em does}
increment $R_i$.  In the third case we can never add a new transition
because, if $R_i$ is not decremented, then it cannot become zero and
the conditions of $q\to q_{succ}$ cannot be satisfied.  On the other
hand, if $q_{pred}\to q$ decrements $R_i$, $q$ would not be
compressible.  Finally, in the fourth case, we should always add the
new transition because, by the supposition that $q$ is compressible,
we know that $q_{pred}\to q$ does not decrement $R_i$ and there is no
chance that it becomes empty.

Now, the state reduction algorithm is defined as iterative reduction of
compressible states.  Using this algorithm, it is possible to compress the
graph corresponding to the original register machine $U_{22}$ to a construct
with 7 states, including a $STOP$ state.  We will refer to this construct as $\mathcal{U}_7$;
the program of $\mathcal{U}_7$ can be found in the appendix.
The Petri net associated with $\mathcal{U}_7$ has 14 places, 31
transitions, 51 inhibitor arcs, and the maximal transition degree equal to 8;
the descriptional complexity of this net is therefore $(14, 31, 51, 8)$.
At the initial marking, the place \place{Q3}, obtained by compression of state $q_3$,
will contain one token.
We will refer to this net as $N_2$.

\subsection{Binary Coding of State Numbers}\label{ssec:binary}
We can further reduce the number of places of the universal Petri net by
avoiding the allocation of a place per state, but instead coding the current
state number in binary.  If the simulated register machine has $n$ states, we
will use $\lceil \log_2 n\rceil = n_p$ places to codify the current state
number in the following way: the place \place{Qi}, $0\leq i < n_p$, contains a
token if the $i$-th bit of the binary representation of $n$ is one, and is
empty otherwise.  All transitions of such a Petri net will thus depend on all
the state places \place{Qi}, $0\leq i < n_p$, and will produce the new
marking of the state places corresponding to the next state number.

\begin{figure}
  \centering
  \includegraphics[scale=0.5]{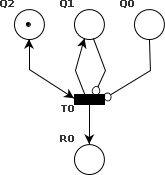}
  \caption{A Petri net simulating a $RiP$ instructions with binary
    coded states.}
  \label{fig:RiPBin}
\end{figure}

As an example, consider the following instruction of an imaginary register
machine with 8 states: $(q_4, R0P, q_6)$.  This instruction defines one
transition going out of the state $q_4$ and into state $q_6$; we will need
$\lceil \log_2 8 \rceil = 3$ state places to simulate this
register machine.
Supposing that the numbering of states is zero-based, the binary code for state
$q_4$ will be $(100)_2$, and $(110)_2$ for $q_6$.
Therefore
 we can draw the Petri net simulating this transition with binary-coded state
numbers as shown in Figure~\ref{fig:RiPBin}.
The transition can only be fired when \place{Q2} contains at least a token, and
\place{Q1} and \place{Q0} are empty, which corresponds to the binary number
$(100)_2$ and to state $q_4$.  The effect of this transition is adding a token
to the register place \place{R0}, leaving the token of \place{Q2} in place,
putting a token into \place{Q1}, and leaving \place{Q0} empty.  The marking of
the state places \place{Q2}, \place{Q1}, and \place{Q0} after
the firing of this transition
corresponds to the binary number $(110)_2$ and to state $q_6$.

Clearly, coding states in binary can be done for any register machine. We will
apply this technique to the construct $\mathcal{U}_7$ obtained previously.
That machine had 7 states, which means that we will need
3 state places with binary coding in the associated net.  This amounts to $11$ places all in all.
Because of the encoding more arcs should be added to each transition and this
augments the maximal transition degree to the value 11.
As far as the number of inhibitor arcs is concerned, it is important to realize
that this parameter will also vary depending on how exactly the states are
numbered, because the number of transitions going out of each state differs.
The strategy we adopt in order to keep the number of inhibitors slightly lower
is to assign the greater binary number to the state with more outgoing arcs.
Since checking for a zero binary digit takes an inhibitor arc, following this
strategy will minimize the number of such arcs in the net. Moreover, it is
possible to eliminate the STOP state by erasing corresponding incoming
transitions, hence yielding the net into a deadlock (which corresponds to the
halting condition for the Petri nets).
  The number we have
obtained is 28 extra arcs for
reading the binary-coded state,
which gives 79 inhibitor arcs
all in all, including the 51 arcs resulting from state compression.

Finally, the number of transitions of the Petri net employing this
binary coding of states is the same as that of the
net $N_2$: 31. This amounts to the following possible descriptional
complexity: $(11, 31, 79, 11)$.  We will refer to this net as $N_3$.
The construct $\mathcal{U}_7$ starts in state 3 which is assigned the
code $(010)_2$, which means that, at the initial marking, $N_3$ will
contain one token in place \place{Q1}.

\section{Universal Net with a Small Number of Transitions}\label{sec:mint}

We consider the construction given in the Section~\ref{sec:mind} and we show
how the number of transitions can be decreased. We use the following
observation already formulated in Subsection~\ref{ssec:state compression}: the
increment instructions of the register machine $U_{22}$ can be simulated during
previous zero check and decrement instruction, so the corresponding state can
be omitted from the net, see Figure~\ref{fig:RiZMRjP}. Applied to the
net $N_1$ this process eliminates 9 states and 9
transitions, yielding a new universal net $N_4$ having the descriptional complexity described by
the vector $(21, 25, 13, 5)$.  At the initial marking, $N_4$ will have
one token in place \place{Q1} corresponding to state $q_1$.

\section{Universal Net with a Small Number of Inhibitor Arcs}\label{sec:mini}
We have seen that Petri nets we have constructed so far used inhibitor arcs
heavily.  From the Petri nets we constructed, those having the smallest number
of inhibitor arcs is $N_1$ with 13 inhibitor arcs. It turns out that it is
possible to almost halve this parameter and reduce the number of inhibitor arcs
to 8 -- one per each register of the simulated register machine. The main idea
is to centralize the procedure of checking whether a register is zero and to
reuse the checker subnet whenever this kind of information is required.

\begin{figure}[hb]
  \centering
  \includegraphics[scale=0.5]{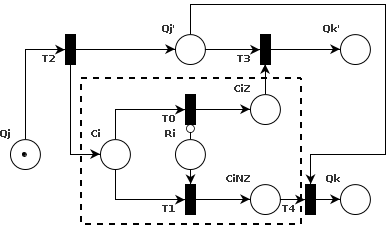}
  \caption{A Petri net simulating a $RiZM$ instruction using a checker subnet.}
  \label{fig:RiZMCheckers}
\end{figure}

Consider the following $RiZM$ instruction: $(q_j, RiZM, q_k, q_{k'})$.
Figure~\ref{fig:RiZMCheckers} shows a Petri net which simulates this instruction.  This
net works in the following way.  Whenever it is required to take a decision
based on the contents of the register $R_i$, the state token from \place{Qj}
is ``split'' into two: one token goes into the waiting place \place{Qj'} and
another token goes into $C_i$ of the $R_i$ checker block (highlighted in the figure).
The checker block checks the value of the register (performs the actual $RiZM$
instruction), and moves the token from \place{Ci} to \place{CiZ} or
\place{CiNZ}, depending on the value of $Ri$.  Whenever a token is put into
\place{CiZ} or \place{CiNZ}, either the transition \trans{T3} or \trans{T4} is
activated, moving the state token into \place{Qk'} if $Ri$ is zero or into
\place{Qk} if the value of the register was nonzero and has thus been
decremented.

Clearly, the checker block can be shared across as many simulations of $RiZM$
instructions as necessary, which means that indeed the number of inhibitor arcs
can be reduced to 8.

We apply the idea of register checkers to the Petri net $N_1$. To count the states, we
note that each checker subnet contains three places plus one for the register,
which makes it 32 places for registers and checkers. The simulation of a $RiZM$
requires one extra place in the model with checkers, which amounts to 13 extra
places, one per each of the
13 $RiZM$ instructions. Finally, there are 22 states, which translates to 22
places.  All in all, the Petri net will contain $32 + 13 + 22 = 67$ places.

As far as the maximal degree of a transition is concerned, note that we do not
use transitions of degree greater than 3 in the simulation of any
instruction. This brings focus upon the clear
trade-off between the number of states and the number of inhibitor arcs.

The number of transitions
%
of the Petri net using checker blocks can
be computed as follows.  Each checker block introduces two new
transitions, which amounts to 16 transitions for eight checker blocks.
Further, we need three transitions per each simulation of a $RiZM$
instructions, which amounts to 39 transitions for the 13 such
instructions of $U_{22}$.  Finally, there are 9 $RiP$ instructions and
9 Petri net transitions to simulate them.  All in all, there are 64
transitions.  The descriptional complexity of this net called $N_5$ is therefore
$(67, 64, 8, 3)$.

By attaching $RiP$ instructions to the previous decrement instructions, the
number of states and transitions can be slightly decreased; the price to be paid
is the increase in the maximal degree. The corresponding descriptional
complexity is $(58, 55, 8, 5)$.  We will refer to this net as $N_6$.

Both $N_5$ and $N_6$ will contain a token in place \place{Q1} at their
initial markings.

\section{Main Results}
To summarize the results we have obtained in the paper, we formulate the
following statement.
\begin{theorem}\label{thm:strong}
  There exist strongly universal Petri nets of following sizes:\\
  $(30,34,13,3)$, $(14, 31, 51, 8)$, $(11, 31, 79, 11)$, $(21,25,13,5)$,
  $(67, 64, 8, 3)$, and $(58, 55, 8, 5)$.
\end{theorem}

In a completely analogous way, the ideas shown previously can be applied to the
weakly universal register machine $U_{20}$ from point (b2) of the Main Theorem
from~\cite{Korec}.  We thus formulate the following result.
\begin{theorem}
  There exist weakly universal Petri nets of following sizes:\\
  $(27, 31, 12, 3)$, $(14,21,23,8)$, $(10,21,44,10)$, $(20,24,12,5)$,
  $(64, 60, 8, 3)$, and $(57,53,8,3)$.
\end{theorem}

\section{Conclusion}

In this paper we constructed 6 strongly universal Petri nets with inhibitor
arcs with sizes shown in Theorem~\ref{thm:strong},
trying to propose an optimal result for each parameter.
While the parameters' space is 4-dimensional, our results
particularly exhibit pairwise relations between some parameters. As a future
work it could be interesting to consider constructions where the trade-off
between 3 or 4 parameters can be observed.

Another interesting question is the minimal value for each of the parameters.
It is clear that the minimal value for the degree is equal to 3 thus we reached
the lower bound in this direction. The number of needed inhibitor arcs is at
least two, as nets with one arc have a decidable reachability
problem~\cite{Reinhardt08,BLM89} which implies that in the computational
variant the halting is decidable.
Hence there cannot be universal nets with one inhibitor arc
for the
class of partially recursive functions. So it could be an interesting challenge
to construct a small net having 2 or 3 inhibitor arcs only.

For the other parameters we cannot indicate a lower bound; we conjecture that
the number of transitions cannot be below 25 (which is the number of branching
points in the Korec machine). We also think that it would be difficult to
substantially decrease the number of states with respect to what was obtained
in the paper.

We would also like to remark that since the original register machine is
deterministic and since our translations preserve this property, the resulting
Petri nets are also deterministic.




\bibliography{sunivpn}

\begin{thebibliography}{10}

\bibitem{AV11}
Artiom Alhazov and Sergey Verlan.
\newblock Minimization strategies for maximally parallel multiset rewriting
  systems.
\newblock {\em Theoretical Computer Science}, 412(17):1581 -- 1591, 2011.

\bibitem{Barzdin}
Ian~M. Barzdin.
\newblock Ob odnom klasse machin {T}uringa (machiny {M}inskogo), russian.
\newblock {\em Algebra i Logika}, 1:42--51, 1963.

\bibitem{BLM89}
Hans~Kleine B{\"u}ning, Theodor Lettmann, and Ernst~W. Mayr.
\newblock Projections of vector addition system reachability sets are
  semilinear.
\newblock {\em Theoretical Computer Science}, 64(3):343--350, 1989.

\bibitem{Korec}
Ivan Korec.
\newblock Small universal register machines.
\newblock {\em Theoretical Computer Science}, 168(2):267--301, 1996.

\bibitem{Malcev}
Anatoly~I. Malcev.
\newblock {\em Algorithms and Recursive Functions}.
\newblock Groningen, Wolters-Noordhoff Pub. Co., 1970.

\bibitem{Minsky62}
Marvin Minsky.
\newblock Size and structure of universal {T}uring machines using tag systems.
\newblock In {\em Recursive Function Theory: Proceedings, Symposium in Pure
  Mathematics, Provelence}, volume~5, pages 229--238, 1962.

\bibitem{Minsky67}
Marvin Minsky.
\newblock {\em Computations: Finite and Infinite Machines}.
\newblock Prentice Hall, Englewood Cliffts, NJ, 1967.

\bibitem{NW12}
Turlough Neary and Damien Woods.
\newblock The complexity of small universal {T}uring machines: A survey.
\newblock In M{\'a}ria Bielikov{\'a}, Gerhard Friedrich, Georg Gottlob, Stefan
  Katzenbeisser, and Gy{\"o}rgy Tur{\'a}n, editors, {\em SOFSEM 2012: 38th
  Conference on Current Trends in Theory and Practice of Computer Science},
  volume 7147 of {\em Lecture Notes in Computer Science}, pages 385--405.
  Springer, 2012.

\bibitem{Ollinger02}
Nicolas Ollinger.
\newblock The quest for small universal cellular automata.
\newblock In Peter Widmayer, Stephan Eidenbenz, Francisco Triguero, Rafael
  Morales, Ricardo Conejo, and Matthew Hennessy, editors, {\em Automata,
  Languages and Programming, 29th International Colloquium, ICALP 2002}, volume
  2380 of {\em Lecture Notes in Computer Science}, pages 318--329. Springer,
  2002.

\bibitem{detpelz}
Elisabeth Pelz.
\newblock Closure properties of deterministic {P}etri nets.
\newblock In {\em Symposium on Theoretical Aspects of Computer Science, STACS
  '87}, volume 247 of {\em Lecture Notes in Computer Science}, pages 371--382.
  Springer, 1986.

\bibitem{Reinhardt08}
Klaus Reinhardt.
\newblock Reachability in {P}etri nets with inhibitor arcs.
\newblock {\em Electronic Notes in Theoretical Computer Science}, 223:239--264,
  2008.

\bibitem{Rogozhin96}
Yurii Rogozhin.
\newblock Small universal {T}uring machines.
\newblock {\em Theoretical Computer Science}, 168(2):215--240, 1996.

\bibitem{handbook}
Grzegorz Rozenberg and Arto Salomaa, editors.
\newblock {\em Handbook of Formal Languages}, volume 1--3.
\newblock Springer, 1997.

\bibitem{Shanon}
Claude~E. Shannon.
\newblock A universal {T}uring machine with two internal states.
\newblock {\em Automata Studies, Annals of Mathematics Studies}, 34:157--165,
  1956.

\bibitem{turing1936}
Alan~M. Turing.
\newblock On computable numbers, with an application to the
  {E}ntscheidungsproblem.
\newblock {\em Proceedings of the London Mathematical Society}, 42(2):230--265,
  1936.

\bibitem{vNeumann66}
John von Neumann.
\newblock {\em Theory of Self-Reproducing Automata}.
\newblock University of Illinois Press, 1966.

\bibitem{Watanabe61}
Shinichi Watanabe.
\newblock 5-symbol 8-state and 5-symbol 6-state universal {T}uring machines.
\newblock {\em Journal of the ACM}, 8(4):476--483, 1961.

\bibitem{Wolfram02}
Stephen Wolfram.
\newblock {\em A New Kind of Science}.
\newblock Wolfram Media Inc., 2002.

\bibitem{Zaitsev2012}
Dmitry~A. Zaitsev.
\newblock Universal {P}etri net.
\newblock {\em Cybernetics and Systems Analysis}, 48(4):498--511, 2012.

\bibitem{Zaitsev13}
Dmitry~A. Zaitsev.
\newblock A small universal {P}etri net.
\newblock {\em EPTCS}, 128:190--202, 2013.
\newblock In Proceedings of Machines, Computations and Universality (MCU 2013),
  arXiv:1309.1043.

\end{thebibliography}



\newpage
\appendix
\section{Descriptions of Some of the Constructions}
In this appendix, we will give the incidence matrices of some of the
universal Petri nets constructed in this paper.  The columns of these
tables correspond to states, the rows to transitions, and each cell of
represents the pair of arc weights $W(p,t), W(t,p)$.

\begin{table}[h]
  \centering
  {\setlength{\tabcolsep}{5.5pt}
\tiny{}\begin{tabular}{l|c|c|c|c|c|c||c|c|c|c|c|c|c|c}&Q3&Q4&Q10&Q16&Q18&Q20&R0&R1&R2&R3&R4&R5&R6&R7\\\hline T1&1,1&&&&&&&1,0&&&&&&0,1\\T2&1,0&0,1&&&&&&-1,0&&&&&0,1&0,1\\T3&&1,1&&&&&&&&&&-1,0&-1,0&\\T4&&1,1&&&&&&&&&&1,0&0,1&\\T5&&1,0&0,1&&&&&&&&&-1,1&1,0&\\T6&0,1&&1,0&&&&&1,0&&&&&-1,0&-1,0\\T7&0,1&&1,0&&&&&1,0&&&1,0&&2,0&-1,0\\T8&&0,1&1,0&&&&&0,1&&&&&-1,0&1,0\\T9&&0,1&1,0&&&&&-1,0&&&1,0&&2,1&-1,0\\T10&&0,1&1,0&&&&&-1,0&&&&&-1,1&-1,0\\T11&&&1,1&&&&&0,1&&&&0,1&1,0&1,0\\T12&&&1,0&0,1&&&&&&&-1,0&&2,0&-1,0\\T13&0,1&&&1,0&&&1,0&1,0&-1,0&&&-1,0&&\\T14&0,1&&&1,0&&&&1,0&1,0&&1,0&-1,0&&\\T15&0,1&&&1,0&&&-1,0&1,0&-1,0&&1,0&-1,0&&\\T16&&0,1&&1,0&&&1,0&-1,0&-1,0&&&-1,0&0,1&\\T17&&0,1&&1,0&&&&-1,0&1,0&&1,0&-1,0&0,1&\\T18&&0,1&&1,0&&&-1,0&-1,0&-1,0&&1,0&-1,0&0,1&\\T19&&&&1,0&0,1&&&&&&&1,0&&\\T20&&&&1,0&&&-1,0&&-1,0&&-1,0&-1,0&&\\T21&&&&1,0&&&&&1,0&&-1,0&-1,0&&\\T22&0,1&&&&1,0&&0,1&1,0&&-1,0&&-1,0&&\\T23&0,1&&&&1,0&&&1,0&&1,0&1,0&-1,0&&\\T24&&0,1&&&1,0&&0,1&-1,0&&-1,0&&-1,0&0,1&\\T25&&0,1&&&1,0&&&-1,0&&1,0&1,0&-1,0&0,1&\\T26&&&&&1,0&0,1&&&&&&1,0&&\\T27&&&&&1,0&&&&&1,0&-1,0&-1,0&&\\T28&0,1&&&&&1,0&&1,0&0,1&0,1&1,0&-1,0&&\\T29&&0,1&&&&1,0&&-1,0&0,1&0,1&1,0&-1,0&0,1&\\T30&&&&0,1&&1,0&&&&&0,1&1,0&&\\T31&&&&&&1,0&&&0,1&0,1&-1,0&-1,0&&\\\end{tabular}}
  \caption{The incidence table of the Petri net $N_2$ of size $(14, 31, 51, 8)$.}
\end{table}

\begin{table}
  \centering
  {\tiny{}\begin{tabular}{l|c|c|c||c|c|c|c|c|c|c|c}&Q0&Q1&Q2&R0&R1&R2&R3&R4&R5&R6&R7\\\hline T1&-1,0&1,1&-1,0&&1,0&&&&&&0,1\\T2&-1,1&1,1&-1,0&&-1,0&&&&&0,1&0,1\\T3&1,1&1,1&-1,0&&&&&&-1,0&-1,0&\\T4&1,1&1,1&-1,0&&&&&&1,0&0,1&\\T5&1,0&1,1&-1,1&&&&&&-1,1&1,0&\\T6&-1,0&1,1&1,0&&1,0&&&&&-1,0&-1,0\\T7&-1,0&1,1&1,0&&1,0&&&1,0&&2,0&-1,0\\T8&-1,1&1,1&1,0&&0,1&&&&&-1,0&1,0\\T9&-1,1&1,1&1,0&&-1,0&&&1,0&&2,1&-1,0\\T10&-1,1&1,1&1,0&&-1,0&&&&&-1,1&-1,0\\T11&-1,0&1,1&1,1&&0,1&&&&0,1&1,0&1,0\\T12&-1,1&1,1&1,1&&&&&-1,0&&2,0&-1,0\\T13&1,0&1,1&1,0&1,0&1,0&-1,0&&&-1,0&&\\T14&1,0&1,1&1,0&&1,0&1,0&&1,0&-1,0&&\\T15&1,0&1,1&1,0&-1,0&1,0&-1,0&&1,0&-1,0&&\\T16&1,1&1,1&1,0&1,0&-1,0&-1,0&&&-1,0&0,1&\\T17&1,1&1,1&1,0&&-1,0&1,0&&1,0&-1,0&0,1&\\T18&1,1&1,1&1,0&-1,0&-1,0&-1,0&&1,0&-1,0&0,1&\\T19&1,1&1,0&1,1&&&&&&1,0&&\\T20&1,0&1,0&1,0&-1,0&&-1,0&&-1,0&-1,0&&\\T21&1,0&1,0&1,0&&&1,0&&-1,0&-1,0&&\\T22&1,0&-1,1&1,0&0,1&1,0&&-1,0&&-1,0&&\\T23&1,0&-1,1&1,0&&1,0&&1,0&1,0&-1,0&&\\T24&1,1&-1,1&1,0&0,1&-1,0&&-1,0&&-1,0&0,1&\\T25&1,1&-1,1&1,0&&-1,0&&1,0&1,0&-1,0&0,1&\\T26&1,0&-1,0&1,1&&&&&&1,0&&\\T27&1,0&-1,0&1,0&&&&1,0&-1,0&-1,0&&\\T28&-1,0&-1,1&1,0&&1,0&0,1&0,1&1,0&-1,0&&\\T29&-1,1&-1,1&1,0&&-1,0&0,1&0,1&1,0&-1,0&0,1&\\T30&-1,1&-1,1&1,1&&&&&0,1&1,0&&\\T31&-1,0&-1,0&1,0&&&0,1&0,1&-1,0&-1,0&&\\\end{tabular}}
  \caption{The incidence table of a Petri net $N_3$ of size $(11, 31, 79, 11)$.}
\end{table}

\begin{table}
\centering
{\tiny{}
\begin{turn}{-90}
\setlength{\tabcolsep}{3pt}
\begin{tabular}{l|c|c|c|c|c|c|c|c|c|c|c|c|c|c|c|c|c|c|c|c|c|c||c|c|c|c|c|c|c|c}&Q1&Q3&Q4&Q6&Q7&Q9&Q10&Q12&Q13&Q14&Q16&Q18&Q20&Q22&Q23&Q25&Q27&Q29&Q30&Q31&Q32&Q33&R0&R1&R2&R3&R4&R5&R6&R7\\\hline T1&1,0&0,1&&&&&&&&&&&&&&&&&&&&&&1,0&&&&&&\\T2&1,0&&&0,1&&&&&&&&&&&&&&&&&&&&-1,0&&&&&&\\T3&0,1&1,0&&&&&&&&&&&&&&&&&&&&&&&&&&&&0,1\\T4&&&1,0&0,1&&&&&&&&&&&&&&&&&&&&&&&&1,0&&\\T5&&&1,0&&0,1&&&&&&&&&&&&&&&&&&&&&&&-1,0&&\\T6&&&0,1&1,0&&&&&&&&&&&&&&&&&&&&&&&&&0,1&\\T7&&&0,1&&1,0&&&&&&&&&&&&&&&&&&&&&&&&-1,0&\\T8&&&&&1,0&0,1&&&&&&&&&&&&&&&&&&&&&&&1,0&\\T9&&&&&&1,0&0,1&&&&&&&&&&&&&&&&&&&&&0,1&&\\T10&&&&&&&1,0&0,1&&&&&&&&&&&&&&&&&&&&&&1,0\\T11&&&&&&&1,0&&0,1&&&&&&&&&&&&&&&&&&&&&-1,0\\T12&&&&&0,1&&&1,0&&&&&&&&&&&&&&&&0,1&&&&&&\\T13&0,1&&&&&&&&1,0&&&&&&&&&&&&&&&&&&&&-1,0&\\T14&&&&&&&&&1,0&&&&&&&&&&&&&0,1&&&&&&&1,0&\\T15&0,1&&&&&&&&&1,0&&&&&&&&&&&&&&&&&1,0&&&\\T16&&&&&&&&&&1,0&0,1&&&&&&&&&&&&&&&&-1,0&&&\\T17&&&&&&&&&&&1,0&0,1&&&&&&&&&&&&&&&&1,0&&\\T18&&&&&&&&&&&1,0&&&&0,1&&&&&&&&&&&&&-1,0&&\\T19&&&&&&&&&&&&1,0&0,1&&&&&&&&&&&&&&&1,0&&\\T20&&&&&&&&&&&&1,0&&&&&0,1&&&&&&&&&&&-1,0&&\\T21&&&&&&&&&&&&&1,0&0,1&&&&&&&&&&&&&&1,0&&\\T22&&&&&&&&&&&&&1,0&&&&&&0,1&&&&&&&&&-1,0&&\\T23&&&&&&&&&&&0,1&&&1,0&&&&&&&&&&&&&0,1&&&\\T24&&&&&&&&&&&&&&&1,0&0,1&&&&&&&&&-1,0&&&&&\\T25&&&&&&&&&&&&&&&1,0&&&&&&0,1&&&&1,0&&&&&\\T26&0,1&&&&&&&&&&&&&&&1,0&&&&&&&1,0&&&&&&&\\T27&&&&&&&&&&&&&&&&1,0&&&&&0,1&&-1,0&&&&&&&\\T28&&&&&&&&&&&&&&&&&1,0&0,1&&&&&&&&-1,0&&&&\\T29&&&&&&&&&&&&&&&&&1,0&&&&0,1&&&&&1,0&&&&\\T30&0,1&&&&&&&&&&&&&&&&&1,0&&&&&0,1&&&&&&&\\T31&&&&&&&&&&&&&&&&&&&1,0&0,1&&&&&0,1&&&&&\\T32&&&&&&&&&&&&&&&&&&&&1,0&0,1&&&&&0,1&&&&\\T33&0,1&&&&&&&&&&&&&&&&&&&&1,0&&&&&&1,0&&&\\T34&&&&&&&&&&0,1&&&&&&&&&&&&1,0&&&&&&&0,1&\\\end{tabular}
\end{turn}
}
\caption{The incidence table of the Petri net $N_1$ of size $(30, 34, 13, 3)$.}
\end{table}

\begin{table}
\centering
{
\begin{turn}{-90}
\setlength{\tabcolsep}{3pt}
\tiny{}\begin{tabular}{l|c|c|c|c|c|c|c|c|c|c|c|c|c|c|c|c|c|c|c|c||c|c|c|c|c|c|c}&Q1&Q3&Q4&Q6&Q7&Q9&Q10&Q12&Q13&Q14&Q16&Q18&Q20&Q22&Q23&Q27&Q30&Q31&Q32&Q33&R0&R1&R2&R4&R5&R6&R7\\\hline T1&1,0&0,1&&&&&&&&&&&&&&&&&&&&1,0&&&&&\\T2&1,0&&&0,1&&&&&&&&&&&&&&&&&&-1,0&&&&&\\T3&0,1&1,0&&&&&&&&&&&&&&&&&&&&&&&&&0,1\\T4&&&1,0&0,1&&&&&&&&&&&&&&&&&&&&&1,0&&\\T5&&&1,0&&0,1&&&&&&&&&&&&&&&&&&&&-1,0&&\\T6&&&0,1&1,0&&&&&&&&&&&&&&&&&&&&&&0,1&\\T7&&&0,1&&1,0&&&&&&&&&&&&&&&&&&&&&-1,0&\\T8&&&&&1,0&0,1&&&&&&&&&&&&&&&&&&&&1,0&\\T9&&&&&&1,0&0,1&&&&&&&&&&&&&&&&&&0,1&&\\T10&&&&&&&1,0&0,1&&&&&&&&&&&&&&&&&&&1,0\\T11&&&&&&&1,0&&0,1&&&&&&&&&&&&&&&&&&-1,0\\T12&&&&&0,1&&&1,0&&&&&&&&&&&&&&0,1&&&&&\\T13&0,1&&&&&&&&1,0&&&&&&&&&&&&&&&&&-1,0&\\T14&&&&&&&&&1,0&&&&&&&&&&&0,1&&&&&&1,0&\\T15&0,1&&&&&&&&&1,0&&&&&&&&&&&&&&1,0&&&\\T16&&&&&&&&&&1,0&0,1&&&&&&&&&&&&&-1,0&&&\\T17&&&&&&&&&&&1,0&0,1&&&&&&&&&&&&&1,0&&\\T18&&&&&&&&&&&1,0&&&&0,1&&&&&&&&&&-1,0&&\\T19&&&&&&&&&&&&1,0&0,1&&&&&&&&&&&&1,0&&\\T20&&&&&&&&&&&&1,0&&&&0,1&&&&&&&&&-1,0&&\\T21&&&&&&&&&&&&&1,0&0,1&&&&&&&&&&&1,0&&\\T22&&&&&&&&&&&&&1,0&&&&0,1&&&&&&&&-1,0&&\\T23&&&&&&&&&&&0,1&&&1,0&&&&&&&&&&0,1&&&\\T24&0,1&&&&&&&&&&&&&&1,0&&&&&&-1,0&&&&&&\\T25&&&&&&&&&&&&&&&1,0&&&&0,1&&1,0&&&&&&\\T26&0,1&&&&&&&&&&&&&&&1,0&&&&&&&-1,0&&&&\\T27&&&&&&&&&&&&&&&&1,0&&&0,1&&&&1,0&&&&\\T28&&&&&&&&&&&&&&&&&1,0&0,1&&&0,1&&&&&&\\T29&&&&&&&&&&&&&&&&&&1,0&0,1&&&&0,1&&&&\\T30&0,1&&&&&&&&&&&&&&&&&&1,0&&&&&1,0&&&\\T31&&&&&&&&&&0,1&&&&&&&&&&1,0&&&&&&0,1&\\\end{tabular}
\end{turn}
}
\caption{The incidence table of the Petri net $N^w_1$ of size $(27, 31, 12, 3)$.}
\end{table}

\begin{table}
  \centering
  {\setlength{\tabcolsep}{5.5pt}
\tiny{}
\begin{tabular}{l|c|c|c|c|c|c||c|c|c|c|c|c|c}&Q1&Q4&Q10&Q16&Q18&Q20&R0&R1&R2&R4&R5&R6&R7\\\hline T1&1,1&&&&&&&1,0&&&&&0,1\\T2&1,0&0,1&&&&&&-1,0&&&&0,1&\\T3&&1,1&&&&&&&&&-1,0&-1,0&\\T4&&1,1&&&&&&&&&1,0&0,1&\\T5&&1,0&0,1&&&&&&&&-1,1&1,0&\\T6&0,1&&1,0&&&&&&&&&-1,0&-1,0\\T7&0,1&&1,0&&&&&&&1,0&&1,1&-1,0\\T8&&0,1&1,0&&&&&0,1&&&&-1,0&1,0\\T9&&&1,1&&&&&0,1&&&0,1&1,0&1,0\\T10&&&1,0&0,1&&&&&&-1,0&&1,1&-1,0\\T11&0,1&&&1,0&&&-1,0&&&&-1,0&&\\T12&0,1&&&1,0&&&1,0&&1,0&2,0&-1,0&&\\T13&&&&1,0&0,1&&&&&&1,0&&\\T14&&&&1,0&&&1,0&&&-1,0&-1,0&&\\T15&0,1&&&&1,0&&&&-1,0&&-1,0&&\\T16&0,1&&&&1,0&&&&2,0&2,0&-1,0&&\\T17&&&&&1,0&0,1&&&&&1,0&&\\T18&&&&&1,0&&&&1,0&-1,0&-1,0&&\\T19&0,1&&&&&1,0&0,1&&1,1&2,0&-1,0&&\\T20&&&&0,1&&1,0&&&&0,1&1,0&&\\T21&&&&&&1,0&0,1&&0,1&-1,0&-1,0&&\\\end{tabular}}
  \caption{The incidence table of the Petri net $N^w_2$ of size $(14, 21, 23, 8)$.}
\end{table}

\begin{table}
  \centering
{\tiny{}\begin{tabular}{l|c|c|c||c|c|c|c|c|c|c}&Q0&Q1&Q2&R0&R1&R2&R4&R5&R6&R7\\\hline T1&-1,0&1,1&-1,0&&1,0&&&&&0,1\\T2&-1,0&1,0&-1,1&&-1,0&&&&0,1&\\T3&-1,0&-1,0&1,1&&&&&-1,0&-1,0&\\T4&-1,0&-1,0&1,1&&&&&1,0&0,1&\\T5&-1,1&-1,1&1,1&&&&&-1,1&1,0&\\T6&1,0&1,1&1,0&&&&&&-1,0&-1,0\\T7&1,0&1,1&1,0&&&&1,0&&1,1&-1,0\\T8&1,0&1,0&1,1&&0,1&&&&-1,0&1,0\\T9&1,1&1,1&1,1&&0,1&&&0,1&1,0&1,0\\T10&1,0&1,1&1,1&&&&-1,0&&1,1&-1,0\\T11&-1,0&1,1&1,0&-1,0&&&&-1,0&&\\T12&-1,0&1,1&1,0&1,0&&1,0&2,0&-1,0&&\\T13&-1,1&1,0&1,1&&&&&1,0&&\\T14&-1,0&1,0&1,0&1,0&&&-1,0&-1,0&&\\T15&1,0&-1,1&1,0&&&-1,0&&-1,0&&\\T16&1,0&-1,1&1,0&&&2,0&2,0&-1,0&&\\T17&1,1&-1,1&1,0&&&&&1,0&&\\T18&1,0&-1,0&1,0&&&1,0&-1,0&-1,0&&\\T19&1,0&1,1&-1,0&0,1&&1,1&2,0&-1,0&&\\T20&1,0&1,1&-1,1&&&&0,1&1,0&&\\T21&1,0&1,0&-1,0&0,1&&0,1&-1,0&-1,0&&\\\end{tabular}}
  \caption{The incidence table of a Petri net $N^w_3$ of size $(10,21,44,10)$.}
\end{table}

\begin{table}
  \centering
  \begin{tabular}{r|r||l|l}
    $q_i$ & $q_j$ & Conditions & Operations \\
    \hline
    1 & 1  &  $R_1\neq 0$ & $R1M$, $R7P$ \\
    1 & 2  &  $R_1=0$ & $R6P$, $R7P$ \\
    2 & 2  &  $R_5=0$, $R_6=0$ & \\
    2 & 2  &  $R_5 \neq 0$ & $R5M$, $R6P$ \\
    2 & 3  &  $R_5=0$, $R_6 \neq 0$ & $R5P$, $R6M$ \\
    3 & 1  &  $R_1\neq 0$, $R_6=0$, $R_7=0$ & $R1M$ \\
    3 & 1  &  $R_1\neq 0$, $R_4\neq 0$, $R_6\neq 0$, $R_7=0$ & $R1M$, $R4M$ \\
    3 & 2  &  $R_6=0$, $R_7\neq 0$ & $R1P$, $R7M$ \\
    3 & 2  &  $R_1=0$, $R_4\neq 0$, $R_6\neq 0$, $R_7=0$ & $R4M$, $R6P$ \\
    3 & 2  &  $R_1=0$, $R_6=0$, $R_7=0$ & $R6P$ \\
    3 & 3  &  $R_6\neq 0$, $R_7\neq 0$ & $R1P$, $R5P$, $R6M$, $R7M$ \\
    3 & 4  &  $R_4=0$, $R_6\neq 0$, $R_7=0$ & \\
    4 & 1  &  $R_0\neq 0$, $R_1\neq 0$, $R_2=0$, $R_5=0$ & $R0M$, $R1M$ \\
    4 & 1  &  $R_1\neq 0$, $R_2\neq 0$, $R_4\neq 0$, $R_5=0$ & $R1M$, $R2M$, $R4M$ \\
    4 & 1  &  $R_0=0$, $R_1\neq 0$, $R_2=0$, $R_4\neq 0$, $R_5=0$ & $R1M$, $R4M$ \\
    4 & 2  &  $R_0\neq 0$, $R_1=0$, $R_2=0$, $R_5=0$ & $R0M$, $R6P$ \\
    4 & 2  &  $R_1=0$, $R_2\neq 0$, $R_4\neq 0$, $R_5=0$ & $R2M$, $R4M$, $R6P$ \\
    4 & 2  &  $R_0=0$, $R_1=0$, $R_2=0$, $R_4\neq 0$, $R_5=0$ & $R4M$, $R6P$ \\
    4 & 5  &  $R_5\neq 0$ & $R5M$ \\
    4 & 7  &  $R_0=0$, $R_2=0$, $R_4=0$, $R_5=0$ & \\
    4 & 7  &  $R_2\neq 0$, $R_4=0$, $R_5=0$ & $R2M$ \\
    5 & 1  &  $R_1\neq 0$, $R_3=0$, $R_5=0$ & $R0P$, $R1M$ \\
    5 & 1  &  $R_1\neq 0$, $R_3\neq 0$, $R_4\neq 0$, $R_5=0$ & $R1M$, $R3M$, $R4M$ \\
    5 & 2  &  $R_1=0$, $R_3=0$, $R_5=0$ & $R0P$, $R6P$ \\
    5 & 2  &  $R_1=0$, $R_3\neq 0$, $R_4\neq 0$, $R_5=0$ & $R3M$, $R4M$, $R6P$ \\
    5 & 6  &  $R_5\neq 0$ & $R5M$ \\
    5 & 7  &  $R_3\neq 0$, $R_4=0$, $R_5=0$ & $R3M$ \\
    6 & 1  &  $R_1\neq 0$, $R_4\neq 0$, $R_5=0$ & $R1M$, $R2P$, $R3P$, $R4M$ \\
    6 & 2  &  $R_1=0$, $R_4\neq 0$, $R_5=0$ & $R2P$, $R3P$, $R4M$, $R6P$ \\
    6 & 4  &  $R_5\neq 0$ & $R4P$, $R5M$ \\
    6 & 7  &  $R_4=0$, $R_5=0$ & $R2P$, $R3P$ \\
  \end{tabular}
  \caption{The program of the universal machine $\mathcal{U}_7$. Each
    line corresponds to a transition from state $q_i$ to state $q_j$,
    checking the conditions listed in the ``Conditions'' column and
    performing those operations on registers which are given in the
    ``Operations'' column.}
\end{table}

\end{document}